\documentclass[fleqn,10pt]{wlscirep}
\usepackage[utf8]{inputenc}
\usepackage[T1]{fontenc}

\usepackage{amsmath, amssymb, bm}
\usepackage{multirow}
\usepackage[subrefformat=parens]{subcaption}

\usepackage{soul}

\def\coloneqq{\mathrel{\mathop:}=}%
\newcommand{\argmin}{\mathop{\rm arg~min}\limits}

\title{Nonnegative/Binary Matrix Factorization for Image Classification using Quantum Annealing}

\author[1*]{Hinako Asaoka}
\author[1,2]{Kazue Kudo}
\affil[1]{Department of Computer Science, Ochanomizu University, Tokyo 112-8610, Japan}
\affil[2]{Graduate School of Information Sciences, Tohoku University, Sendai 980-8579, Japan}

\affil[*]{asaoka.hinako@is.ocha.ac.jp}

\begin{abstract}
Classical computing has borne witness to the development of machine learning. The integration of quantum technology into this mix will lead to unimaginable benefits and be regarded as a giant leap forward in mankind’s ability to compute. Demonstrating the benefits of this integration now becomes essential.
With the advance of quantum computing, several machine-learning techniques have been proposed that use quantum annealing. In this study, we implement a matrix factorization method using quantum annealing for image classification and compare the performance with traditional machine-learning methods. Nonnegative/binary matrix factorization (NBMF) was originally introduced as a generative model, and we propose a multiclass classification model as an application.
We extract the features of handwritten digit images using NBMF and apply them to solve the classification problem. Our findings show that when the amount of data, features, and epochs is small, the accuracy of models trained by NBMF is superior to classical machine-learning methods, such as neural networks. Moreover, we found that training models using a quantum annealing solver significantly reduces computation time.
Under certain conditions, there is a benefit to using quantum annealing technology with machine learning.
\end{abstract}
\begin{document}

\flushbottom
\maketitle
% Click the title above to edit the author's information and abstract
%
\thispagestyle{empty}

% \noindent Please note: Abbreviations should be introduced at the first mention in the main text – no abbreviations lists. The suggested structure of the main text (not enforced) is as follows.

\section*{Introduction}

Quantum computing progresses each year, expanding the scale of problems that quantum computers can solve and improving the accuracy of their solutions. The next stage of development will be to verify the practicality of these advances. To ensure the future spread of quantum technology, it is critical to demonstrate examples in which quantum computers outperform conventional classical computers.

Here, we focus on a machine-learning method that uses a quantum annealing machine. Numerous attempts have been made to apply quantum annealing technology to machine learning, as confirmed by several studies\cite{Willsch_2019, Nath_2021, Barbosa_2021, Yarkoni_2022, Urushibata_2022, Wang_2022, Abel_2022, Ferrari_Dacrema_2022}. These studies demonstrate that quantum annealing effectively reduces data size and learning time, indicating its potential for solving problems that are too challenging for classical machine-learning computation.

Our study concentrates on a method called nonnegative/binary matrix factorization (NBMF). Initially proposed as a generative model trained using a quantum annealing machine, NBMF is an algorithm that extracts features from data and optimizes the combination of features required to reproduce the original data. Previous studies have solved the task of decomposing facial images into feature matrices\cite{NBMF_2018, NBMF_2021}. They demonstrated that optimization using D-Wave\cite{DWave_2011, Gibney2017, DWave_2020}, a quantum-annealing machine, requires less time than classical optimization solvers. However, as a machine-learning method, NBMF is less accurate than its classical counterpart, nonnegative matrix factorization (NMF)\cite{Lee_1999}. Furthermore, no comparisons were made with other machine-learning methods. Classical machine-learning algorithms, such as deep learning, have already been used because of their precision\cite{Rawat_2017}. To recommend the use of NBMF, it is essential to demonstrate the superiority over classical machine-learning algorithms. Previous studies demonstrated data reconstruction using features obtained by NBMF.
However, these previous studies did not demonstrate the effectiveness of the features obtained beyond data reconstruction nor did they evaluate the performance of the method for other machine-learning tasks.

In this study, we propose NBMF as a multiclass image classification model. Image classification is one of the most prominent problems in machine learning with practical applications such as image diagnosis and good classification. We executed the multiclass image classification method using NBMF by decomposing the matrix containing the image data and its class information. Our results demonstrate the application of quantum annealing to image classification tasks. Additionally, we solved classification tasks under the same conditions as classical machine-learning methods, such as neural networks, and compared the results. We identify the aspects of image classification by NBMF that are superior or inferior to classical methods and confirm the advantage of quantum technology over conventional machine-learning methods.

\section*{Models and algorithms}

NBMF\cite{NBMF_2018} extracts features by decomposing data into basis and coefficient matrices. The dataset was converted into a positive matrix to prepare the input for the NBMF method. If the dataset contains $n$-dimensional $m$ data, then the input is an $n \times m$ matrix $V$. The input matrix is decomposed into a base matrix $W$ of $n \times k$, representing the dataset features, and a coefficient matrix $H$ of $k \times m$, representing the combination of features selected to reconstruct the original matrix. Then,
\begin{equation}
 V \approx WH,
  \label{eq:v}
\end{equation}
where $W$ and $H$ are positive and binary matrices, respectively. The column number $k$ of $W$ corresponds to the number of features extracted from the data and can be set to any value.
To minimize the difference between the left and right sides of Eq.~\eqref{eq:v}, $W$ and $H$ are updated alternately as
\begin{eqnarray}
 W & \coloneqq & \argmin_{X \in \mathbb{R}^{+n \times k}} \parallel V - XH \parallel _F + \alpha \parallel X \parallel _F,
  \label{eq:w} \\
 H & \coloneqq & \argmin_{X \in { \{ 0,1 \} }^{k \times m}} \parallel V - WX \parallel _F,
\label{eq:h}
\end{eqnarray}
where $\parallel \cdot \parallel _F$ denotes the Frobenius norm. The components of $W$ and $H$ are initially given randomly. The hyperparameter $\alpha$ is a positive real value that prevents overfitting and is set to $\alpha = 1.0 \times 10^{-4}$.

In previous studies, the Projected Gradient Method (PGM) was used to update Eq.~\eqref{eq:w}\cite{Lin_2007}.
The loss function that updates Eq.~\eqref{eq:w} is defined as
\begin{equation}
 f_{W}(\bm{x}) = \parallel \bm{v} - H^{\mathrm{T}} \bm{x} \parallel^2
  + \alpha \parallel \bm{x} \parallel^2,
\label{eq:W_f}
\end{equation}
where $\bm{x}^{\mathrm{T}}$ and $\bm{v}^{\mathrm{T}}$ are the row vectors of $W$ and $V$, respectively. The gradient of Eq.~\eqref{eq:W_f} is expressed as
\begin{equation}
 \nabla f_{W}
  = -H (\bm{v} - H^T \bm{x})
  + \alpha \bm{x}.
  \label{eq:W_grad}
\end{equation}
The PGM minimizes the loss functions in Eq.~\eqref{eq:W_f} by updating $\bm{x}$:
\begin{equation}
 \bm{x}^{t+1}
  = P[\bm{x}^t - \gamma_t \nabla f_W (\bm{x}^t)],
\label{eq:W_update}
\end{equation} where $\gamma_t$ is the learning rate and
\begin{equation}
 P[x_i] =
  \begin{cases}
  0 & (x_i \leq 0), \\
  x_i & (0 < x_i < x_{\rm{max}}), \\
  x_{\rm{max}} & (x_{\rm{max}} \leq x_i),
  \end{cases}
\label{eq:W_p}
\end{equation}
where $x_{\rm{max}}$ is the upper bound and is set to $x_{\rm{max}}=1$. Eq.~\eqref{eq:W_p} is a projection that keeps the components of $\bm{x}$ nonnegative.

However, because $H$ is a binary matrix, Eq.~\eqref{eq:h} can be regarded as a combinatorial optimization problem that can be minimized by using an annealing method. To solve Eq.~\eqref{eq:h} using a D-Wave machine, a quantum annealing computer, we formulated the loss function as a quadratic unconstrained binary optimization model:
\begin{equation}
 f_{H}(\bm{q}) = \sum_i \sum_r W_{ri}(W_{ri} - 2 v_{r}) q_i + 2 \sum_{i<j} \sum_r W_{ri} W_{rj} q_{i} q_{j},
  \label{eq:H_qubo}
\end{equation}
where $\bm{q}$ and $\bm{v}$ are the column vectors of $H$ and $V$, respectively.

After the alternate updating method converges, we obtain $W$ and $H$ which minimize the difference between the left and right sides of Eq.~\eqref{eq:v}.
$W$ consists of representative features extracted from the input data, and
$H$ represents the combination of features in $W$ using binary values to reconstruct $V$.
Therefore, $V$ can be approximated as the product of $W$ and $H$.

Previous studies used NBMF to extract features from facial images\cite{NBMF_2018}. When the number of annealing steps is small, the computation time is shorter than a classical combinatorial optimization solver. However, using the D-Wave machine is disadvantageous in that the computing time increases linearly with the number of annealings, whereas the classical solver does not significantly change the computing time. The results were compared with NMF\cite{Lee_1999}. Unlike NBMF, matrix $H$ in NMF is positive and not binary. While the matrix $H$ produced by NBMF was sparser than NMF, the difference between $V$ and $WH$ of NBMF was approximately 2.17 times larger than NMF.
Although NBMF can have a shorter data processing time than the classical method, it is inferior to NMF as a machine-learning method in accuracy.
Moreover, because previous studies did not demonstrate tasks beyond data reconstruction, the usefulness of NBMF as a machine-learning model is uncertain.

\begin{figure}[tb]
 \centering
 \includegraphics[scale=0.6]{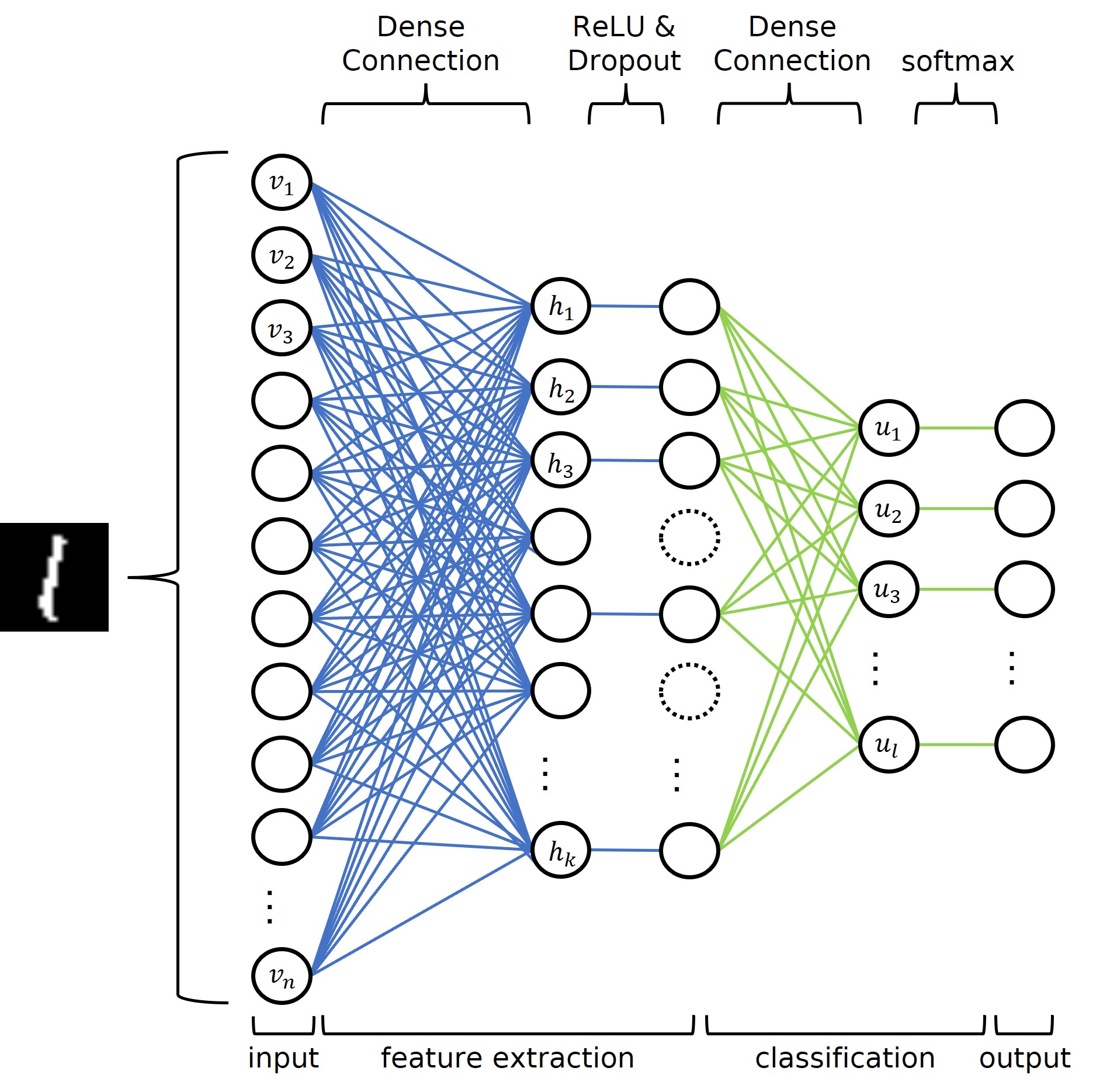}
 \caption{An overview of a fully-connected neural network.}
 \label{fig:FCNN}
\end{figure}

\begin{figure}[tb]
 \centering
 \includegraphics [width=0.5\textwidth]{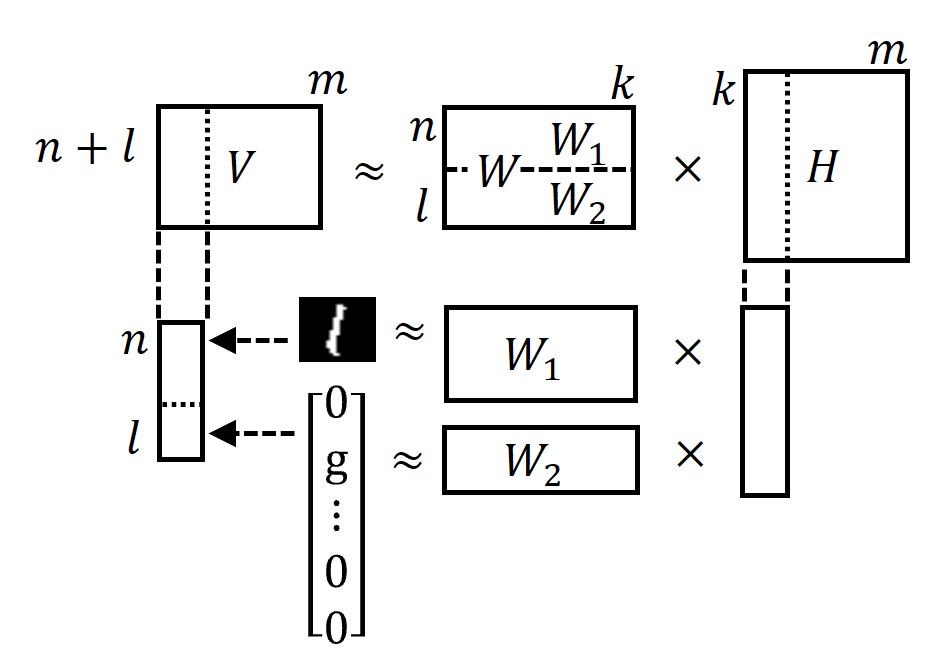}
 \caption{An overview of training by NBMF.}
 \label{fig:NBMF_train}
\end{figure}

In this study, we propose the application of NBMF to a multiclass classification model. Inspired by the structure of a fully connected neural network (FCNN), we define an image classification model using NBMF. In an FCNN, image data are fed into the network as input, as shown in Figure~\ref{fig:FCNN}, and the predicted classes are obtained as the output of the network through the hidden layers.
To perform fully connected network learning using NBMF, we interpret the structure shown in Figure~\ref{fig:FCNN} as a single-matrix decomposition.
When the input and output layers of the FCNN are combined into one input layer, the network becomes a two-layer network with the same structure as NBMF.
As the input to the training network by NBMF, we used a matrix consisting of image data and the corresponding class information. Class information is represented by a one-hot vector multiplied by an arbitrary real number $g$.
The image data and class information vectors are combined row-wise and eventually transformed into an input matrix $V$. We use NBMF to decompose $V$ to obtain the basis matrix $W$ and the coefficient matrix $H$, as shown in Figure~\ref{fig:NBMF_train}.
The column vectors in $H$ correspond to the nodes in the hidden layer of the FCNN network, and the components of $W$ correspond to the weights of the edges.
  The number of feature dimensions $k$ in the NBMF corresponds to the number of nodes in the hidden layer of the FCNN.
To obtain $H$, we minimize Eq.~\eqref{eq:H_qubo} by using an annealing solver, as in a previous study. However, to obtain $W$ by minimizing Eq.~\eqref{eq:W_f}, we propose using the projected Root Mean Square Propagation (RMSProp) method instead of the PGM used in a previous study. RMSProp is a gradient descent method that adjusts the learning and decay rates to help the solution escape local minima\cite{Ruder_2016}. RMSProp updates the vector $\bm{h}$, whose components are denoted by $h_i$ as
\begin{equation}
 h^{t+1}_{i} = \beta h^{t}_{i} + (1-\beta) g^{2}_{i},
  \label{RMSP_h}
\end{equation}
where $\beta$ is the decay rate, $\bm{g} = \nabla f_W$, and vector $\bm{x}$ is
\begin{equation}
 \bm{x}^{t+1} = \bm{x}^{t} - \eta \frac{1}{\sqrt{\bm{h}^{t} + \epsilon}} \nabla f_{W},
  \label{RMSP_x}
\end{equation}
where $\eta$ is the learning rate, and $\epsilon$ is a small value that prevents computational errors. After updating $\bm{x}$ using Eq.~\eqref{RMSP_x}, we apply the projections described in Eq.~\eqref{eq:W_p}, to ensure that the solution does not exceed the bounds. We propose this method as a projected RMSProp.

\begin{figure}[tb]
 \centering
 \includegraphics[scale=0.35]{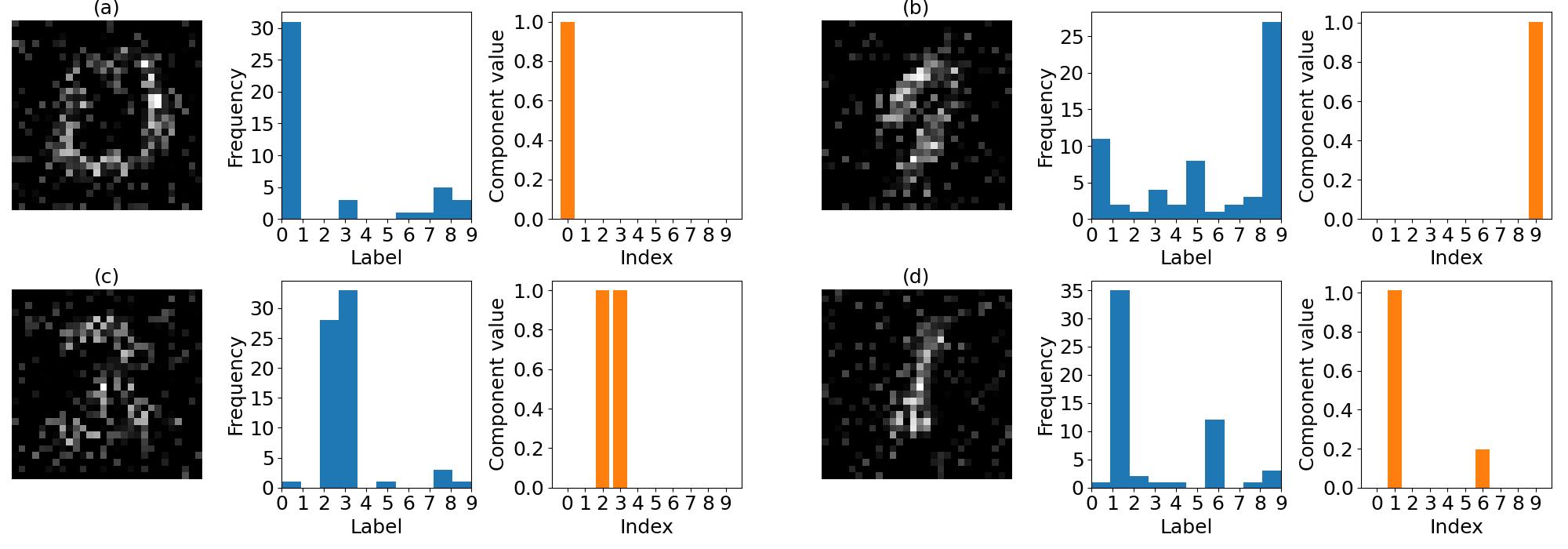}
 \caption{ The figure shows four sets of images, (\textbf{a}), (\textbf{b}), (\textbf{c}), and (\textbf{d}), corresponding to column vectors selected from $W$. Each set contains an image, a histogram, and a bar graph. The image represents a column vector of $W_1$, and the histogram shows how often the column vector was selected to reconstruct the training data images with each label. The orange bar graph plots the component values of the corresponding column vector of $W_2$.}
 \label{fig:W1_W2}
\end{figure}

In Figure~\ref{fig:W1_W2}, we demonstrated the information contained in $W$.
Because the row vectors of $W$ correspond to those of $V$, $W$ consists of $W_1$ corresponding to the image data information, and $W_2$ corresponding to the class information. We plotted four column vectors selected from $W$ trained with MNIST handwritten digit images under the conditions $m = 300$ and $k=40$, as shown in Figure~\ref{fig:W1_W2}. The images in Figure~\ref{fig:W1_W2} show the column vectors of $W_1$. The blue histograms show the frequencies at which the column vectors were selected to reconstruct the training data images with each label. The orange bar graphs show the component values of the corresponding column vectors of $W_2$. For example, the image in Figure~\ref{fig:W1_W2}a resembles Number 0. From the histogram next to the image, we understand that the image is often used to reconstruct the training data labeled as 0. In the bar graph on the right, the corresponding column vector of $W_2$ has the largest component value at an index of 0. This indicates that the column vector corresponding to the image has a feature of Number 0. Similarly, the image in Figure~\ref{fig:W1_W2}b has a label of 9. However, the image in Figure~\ref{fig:W1_W2}c appears to have curved features. From the histogram and bar graph next to the image, it appears that the image is often used to represent labels 2 and 3.
This result is consistent with the fact that both numbers have a curve, which explains why the column vector of $W_1$ was used in the reconstruction of images with labels 2 and 3. The image in Figure~\ref{fig:W1_W2}d has the shape of a straight line, and the corresponding histogram shows that the image is mainly used to express label 1 and is also frequently used to express label 6. Because Number 6 has a straight-line part, the result is reasonable.

\begin{figure}[tb]
 \centering
 \includegraphics[width=0.5\textwidth]{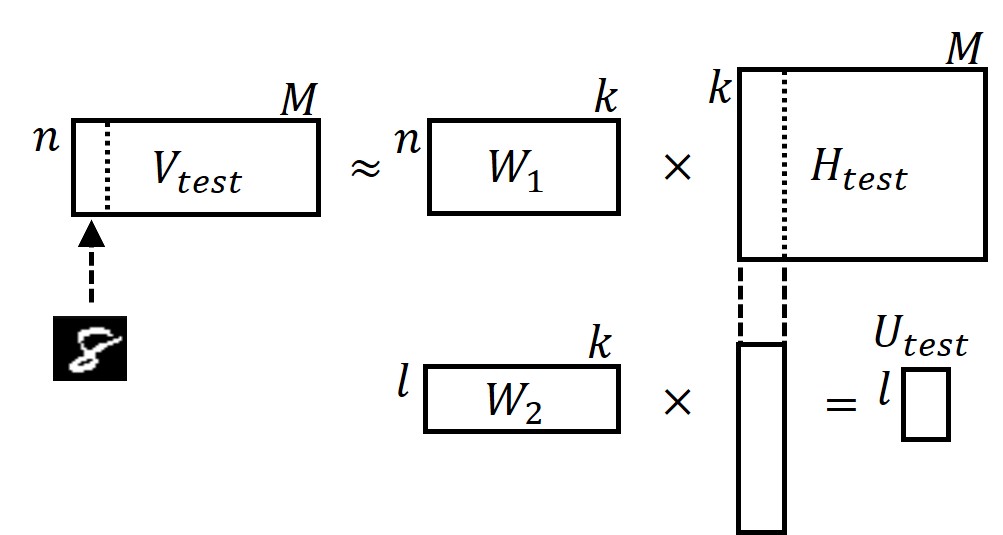}
 \caption{An overview of testing by NBMF.}
 \label{fig:NBMF_test}
\end{figure}

In our multiclass classification model using NBMF, we used the trained matrices $W_1$ and $W_2$ to classify the test data in the workflow shown in Figure \ref{fig:NBMF_test}.
First, we decompose the test data matrix $V_{\rm test}$ to obtain $H_{\rm test}$ by using $W_1$. Here, $M$ represents the amount of test data, which corresponds to the number of column vectors of $V_{\rm test}$. We use Eq.~\eqref{eq:h} for decomposition. Each column vector of $H_{\rm test}$ represents the features selected from the trained $W_1$ to approximate the corresponding column vector of $V_{\rm test}$. Second, we multiply $W_2$ by $H_{\rm test}$ to obtain $U_{\rm test}$, which expresses the prediction of the class vector corresponding to each column vector in $V_{\rm test}$. Finally, we applied the softmax function to the components of $U_{\rm test}$ and considered the index with the largest component value in each column vector to be the predicted class.

\section*{Results}

% Up to three levels of \textbf{subheading} are permitted. The subheadings are not numbered.

\subsection*{High performance for a small number of training data.}

\begin{figure*}[tb]
 \begin{minipage}[b]{0.5\linewidth}
  \centering
  \includegraphics[keepaspectratio,scale=0.4]{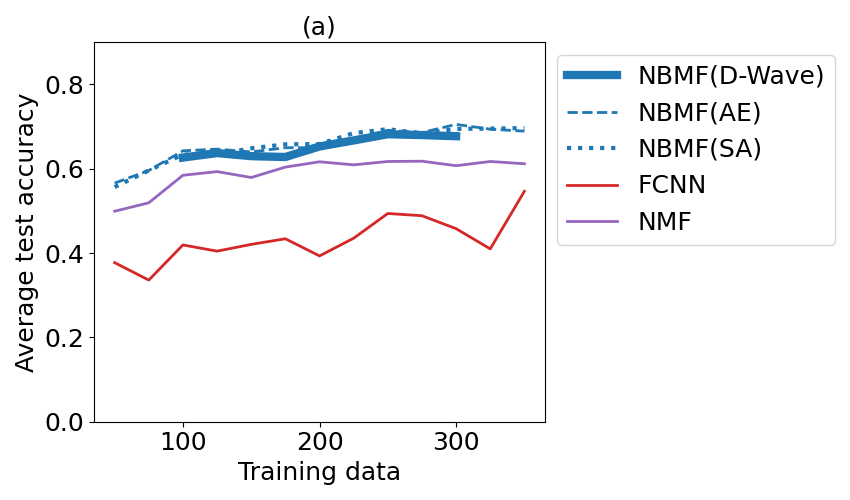}
 \end{minipage}
 \begin{minipage}[b]{0.5\linewidth}
  \centering
  \includegraphics[keepaspectratio,scale=0.4]{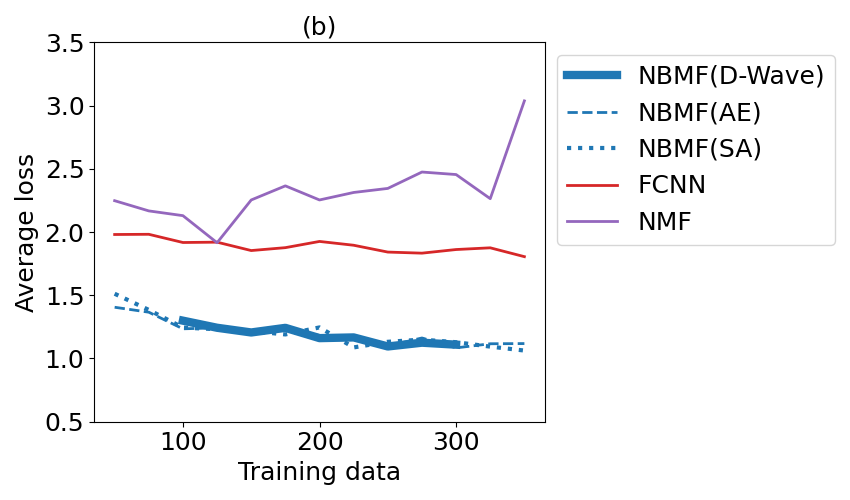}
 \end{minipage}
 \caption{
 The average test accuracy (\textbf{a}) and cross-entropy loss (\textbf{b}) of classification as functions of the amount of training data for each method.
 Averages are taken over three experiments.
 Training methods are NBMF with D-Wave (blue solid), NBMF with Fixstars Amplify annealing engine (AE) (blue dashed), NBMF with simulated annealing (SA) (blue dotted), FCNN (red), and NMF (purple).
 Parameters other than the amount of training data are common in all results.}
 \label{fig:data}
\end{figure*}

We varied the amount of training data $m$ within an arbitrary range and compared the performance of the three machine-learning methods, NBMF, FCNN, and NMF. All methods were tested with only the number of training data varying. Features were fixed at $k=40$, and epochs were limited to ten so that experimental conditions were consistent. In Figure~\ref{fig:data}a, we plot the average accuracy rate as a function of the number of training data and compare the results of the three methods. Our findings reveal that NBMF achieves the highest level of accuracy of classification. Although FCNN and NMF tend to increase in accuracy as the amount of data increases, NBMF can attain a high degree of accuracy even with a small amount of data. In Figure~\ref{fig:data}b, NBMF yields the smallest cross-entropy loss compared with the other methods, even when the amount of training data is small. While the loss of NMF is unstable, NBMF achieves learning stability.

\subsection*{High performance for small feature dimensions.}

\begin{figure*}[tb]
 \begin{minipage}[b]{0.5\linewidth}
  \centering
  \includegraphics[keepaspectratio,scale=0.4]{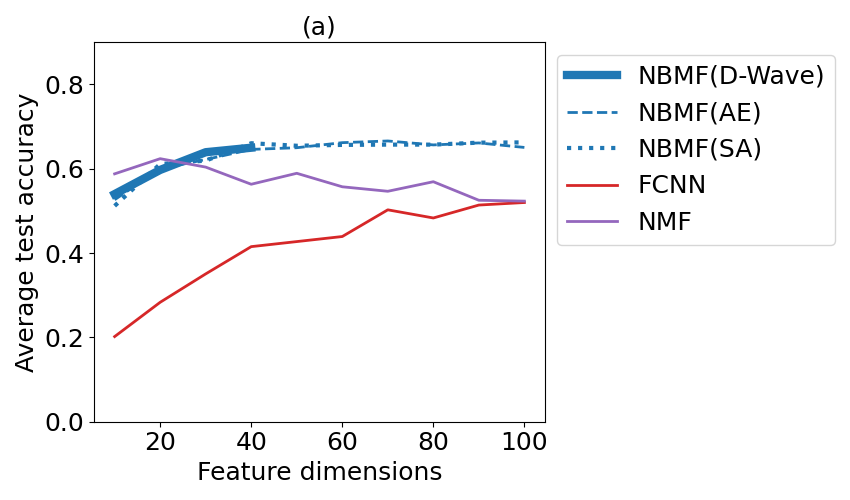}
 \end{minipage}
 \begin{minipage}[b]{0.5\linewidth}
  \centering
  \includegraphics[keepaspectratio,scale=0.4]{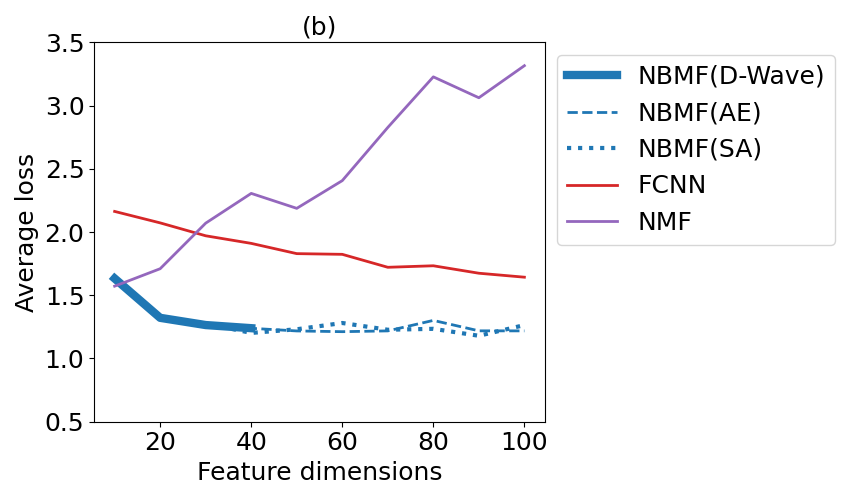}
 \end{minipage}
 \caption{
 The average test accuracy (\textbf{a}) and cross-entropy loss (\textbf{b}) of classification as functions of the number of feature dimensions for each method.
 Averages are taken over three experiments.
 Training methods are NBMF with D-Wave (blue solid), NBMF with Fixstars Amplify AE (blue dashed), NBMF with SA (blue dotted), FCNN (red), and NMF (purple).
 Parameters other than the number of feature dimensions are common in all results.}
 \label{fig:k}
\end{figure*}

We varied the number of features, $k$ within an arbitrary range, and compared the classification performance of the model at an early stage. The number of training data was fixed at $m = 150$. As in the previous experiment, we compared the three machine-learning methods under common conditions. In Figure~\ref{fig:k}a, one significant difference between NBMF and other methods was that NBMF's results were stable even when the number of features was small. While the accuracy of FCNN increases as the number of features increases, and the accuracy of NMF decreases as the number of features increases, NBMF has a stable accuracy rate for the number of features larger than approximately 40. Moreover, in Figure~\ref{fig:k}b, NBMF exhibits the smallest and most stable cross-entropy loss among each of the three methods.

\subsection*{Speeding up using quantum annealing.}

We performed NBMF using three different optimization solvers.
However, as shown in Figures \ref{fig:data} and \ref{fig:k}, the classification performance did not depend significantly on the solvers.
From another perspective, we compare the computation time of each solver in Table \ref{tab:time}.
The total execution time is the shortest for the SA solver because the SA runs locally and does not include communication or queuing times.
However, when comparing the annealing times, the D-Wave machine is the fastest, taking only 1.0 ms.
If the communication and queuing times are not considered, the quantum annealing machine exhibits the best optimization speed.
\begin{table}[tb]
  \caption{
  Comparison of computing time (ms) for three solvers.
  We perform NBMF on a matrix with 10 or 100 columns and measure the average computing time per column.
  Total execution time is measured from when a problem is handed to solvers until when a solution is returned, including communication and queuing time when using cloud services.
  To evaluate pure calculation time, we also obtain annealing time when using a D-Wave machine and the Fixstars Amplify AE.
  Annealing time is included in the total execution time and represents the time that solvers spent on annealing only.}
  \label{tab:time}
  \hbox to\hsize{\hfil
  \scalebox{0.9}{
  \begin{tabular}{ccc||cc}\hline\hline
  & & & \multicolumn{2}{c}{Average time (ms) per vector}\\
  Solver & Usage & Type & $m=10$ & $m=100$ \\\hline
  \multirow{2}{*}{D-Wave} & \multirow{2}{*}{Cloud} & Annealing time & $1.0$ & $1.0$\\
  & & Total execution time & $12.0 \times 10^{3}$ & $8.6 \times 10^{3}$\\
  \multirow{2}{*}{AE} & \multirow{2}{*}{Cloud} & Annealing time & $9.9 \times 10^{3}$ & $9.9 \times 10^{3}$\\
  & & Total execution time & $11.3 \times 10^{3}$ & $10.8 \times 10^{3}$\\
  SA & Local & Total execution time & $0.3 \times 10^{3}$ & $4.2 \times 10^{3}$\\\hline
  \end{tabular}}\hfil}
\end{table}

\section*{Discussion}

% The Discussion should be succinct and must not contain subheadings.

\begin{figure*}[tb]
 \begin{minipage}[b]{0.5\linewidth}
  \centering
  \includegraphics[keepaspectratio,scale=0.4]{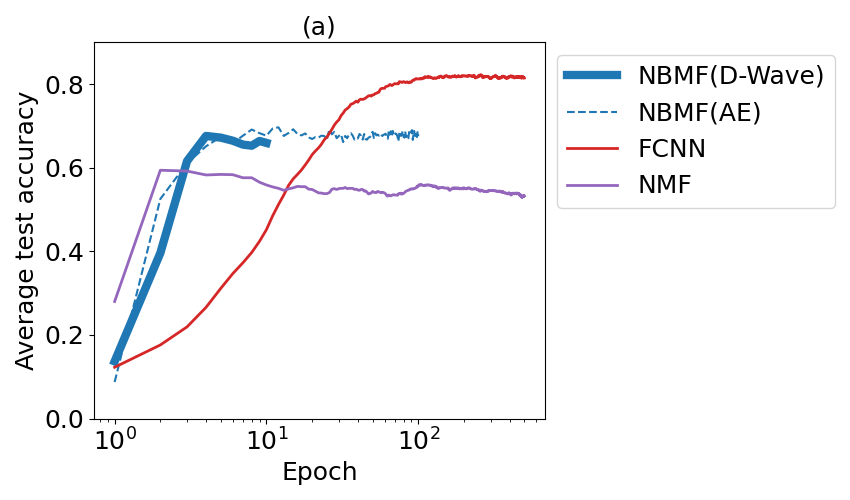}
 \end{minipage}
 \begin{minipage}[b]{0.5\linewidth}
  \centering
  \includegraphics[keepaspectratio,scale=0.4]{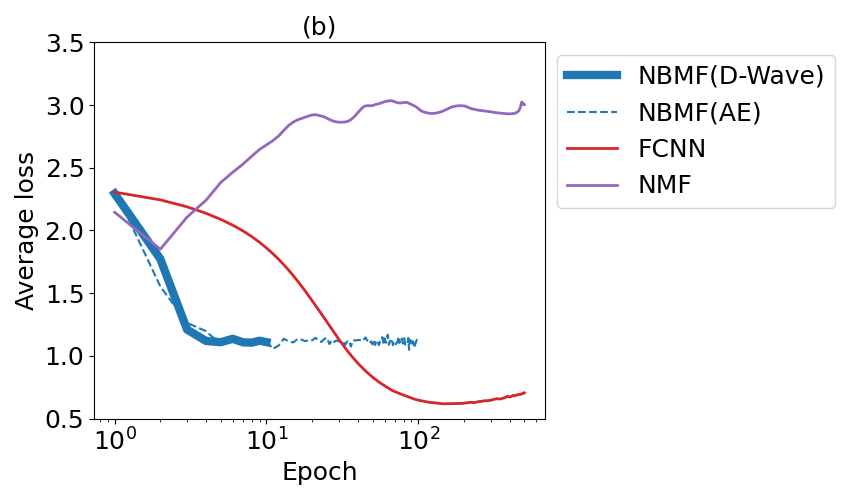}
 \end{minipage}
 \caption{
 Average test accuracy (\textbf{a}) and cross-entropy loss (\textbf{b}) of classification as functions of number of training epochs for each method.
 Averages are taken over three experiments.
 The training methods are NBMF with D-Wave (blue solid), NBMF with Fixstars Amplify AE (blue dashed), NBMF with SA (blue dotted), FCNN (red), and NMF (purple).
 Parameters other than the number of training epochs are common in all results.}
 \label{fig:epoch}
\end{figure*}

We performed image classification using machine learning with NBMF and scored a higher classification accuracy and faster loss reduction than other methods when the amount of training data and number of features were small. However, we found that FCNN outperformed NBMF when trained until there was sufficient convergence. Figure~\ref{fig:epoch} illustrates the rate of accuracy and cross-entropy loss as functions of the number of epochs for the different learning methods. The number of training images and features were fixed at $m = 300$ and $k = 40$, respectively. The results of the NBMF showed that the accuracy rate and loss value were stable from the early stage of approximately ten epochs and did not change significantly afterward. In contrast, the accuracy of FCNN increased as the number of epochs increased, and the results were stable at 100 or more epochs, with the classification accuracy after convergence being higher than NBMF.

The accuracy of NBMF increases rapidly and at an earlier stage than neural networks, although the performance of NBMF is relatively lower than neural networks at a later stage. In conventional machine-learning methods, the accuracy rate usually increases as the amount of training data, features, and epochs increase, as shown in Figures~\ref{fig:data}a, \ref{fig:k}a, and \ref{fig:epoch}a. However, for NBMF, the accuracy rate and error remained stable and showed no significant change, even when the numbers increased. This tendency is because of the difference in the learning approach between NBMF and the other methods. NBMF utilizes a binary combinatorial optimization procedure in the learning process. Previous studies have shown that retaining only essential information using quantum optimization can lead to a higher level of accuracy even when the amount of data is reduced \cite{Mott_2017, Li_2018, Nguyen_2018}.

\begin{figure}[tb]
 \centering
 \includegraphics[scale=0.4]{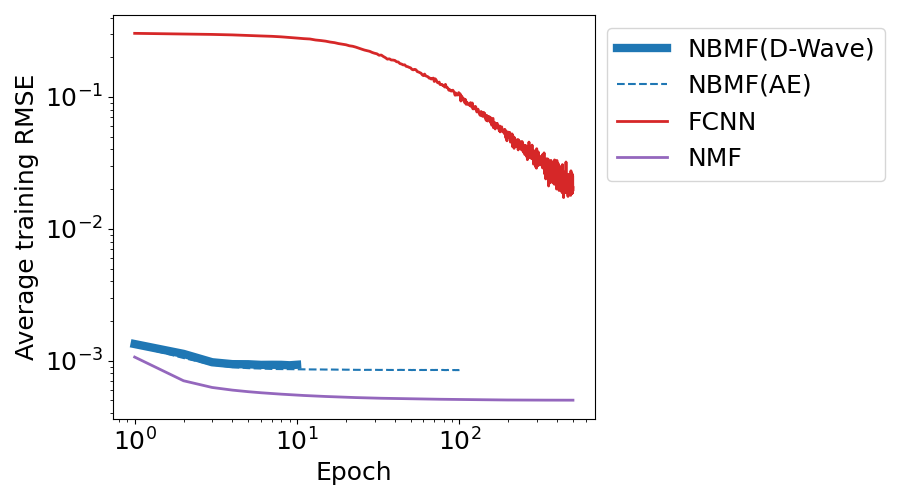}
 \caption{
 Average RMSE of training data as functions of the number of training epochs for each method.
 Averages are taken over three experiments.
 The training methods are NBMF with D-Wave (blue solid), NBMF with Fixstars Amplify AE (blue dashed), FCNN (red), and NMF (purple).
 Parameters other than the number of training epochs are common in all results.}
 \label{fig:rmse}
\end{figure}

We aimed to understand the reason for the poor classification performance of NMF. Figures~\ref{fig:data}b, \ref{fig:k}b, and \ref{fig:epoch}b illustrate that the cross-entropy loss of NMF increases instead of decreasing as the amount of training data, features, and epochs increases, in contrast to NBMF and FCNN. This trend suggests that NMF may be overfitting the training data and may not generalize well to new data.
In Figure~\ref{fig:rmse}, we see that the root mean squared error (RMSE) during training is the lowest among the three methods, indicating that NMF can reproduce the training data well. The poor classification accuracy of NMF suggests that it was too focused on reproducing the training data, causing overfitting.

Table \ref{tab:time} shows the calculation time for optimizing each column in the coefficient matrix $H$ using the different optimization solvers.
We observed that using a D-Wave machine---a quantum annealing computer--- significantly reduced the annealing time.
Reduction in the annealing time will eventually lead to faster machine learning.
The three solvers showed no significant difference in learning performance.
However, the D-Wave machine was significantly faster in terms of annealing time.
We demonstrated the advantage of using a real quantum annealing machine rather than a simulator, which suggests that using a real quantum annealing machine can provide faster computation.

% Conclusion
In this study, we propose applying NBMF, initially presented as a generative model, to a classification model. Previous studies have shown the advantages of using NBMF, such as data compression and high-speed computation, but its usefulness as an applicable machine-learning technique has not been demonstrated.

We solved an image classification task and compared its performance with classical machine-learning methods. Our results demonstrate that learning as a combinatorial optimization problem can achieve higher accuracy than conventional methods at an early stage, even with a small amount of data and few features.
We have also shown that a quantum annealing machine can significantly accelerate computational speed.

Improving the versatility of the classification model is a future direction of this study. While classical machine-learning methods have achieved close to 100\% accuracy in the MNIST digit classification task, deep learning for image recognition is now addressing more challenging and realistic problems, such as automatic driving and medical image diagnosis\cite{Litjens_2017, Caesar_2020, Kadam_2020}.
Our classification method achieved a maximum accuracy of approximately 70\% for MNIST, which is insufficient for solving such problems. One reason for this limited performance is overfitting, as NBMF currently focuses on accurately reproducing the training data. We aim to improve the algorithm to learn more versatile and flexible models that can capture the features of more complex images.

% Revision2
Note the limitations imposed on our experiments are based on the specifications of the quantum computer.
Owing to the limited number of qubits available in the current D-Wave machine, we imposed some limitations on the experiment, such as using a simple grayscale dataset and a limited number of feature dimensions.
Because only Noisy Intermediate-Scale Quantum (NISQ) devices are currently available as practical quantum-computing platforms\mbox{\cite{Preskill_2018}}, these limitations are unavoidable.
Larger-scale problems will be solved more effectively in the post-NISQ era.
To examine the actual performance of machine learning using quantum computers, it is advisable to revisit our experiments after the NISQ era, aligning the experimental conditions with those employed in major deep learning experiments.

Another issue is the communication time with a quantum computer. We highlighted the computing time as an advantage of using a quantum annealing machine for learning. However, communication latency remains a concern. In practice, D-Wave takes the longest time to return a solution among all solvers. Although this is an unavoidable issue with cloud-based machines, the communication time can be shortened depending on the location and network conditions.

We observed that NBMF outperformed classical methods in multiclass classification under certain conditions, for example, when there is a small amount of data or there is a short training time. These results reveal that machine learning by quantum annealing has advantages over classical machine learning methods. We expect that the findings of this study will serve as evidence when practical machine-learning applications using quantum annealing come under consideration.

\section*{Methods}

\subsection*{Classical machine-learning algorithms for comparison.}

To evaluate the performance of NBMF, we compared it with two classical machine-learning methods (FCNN and NMF) by solving the same classification task.
For a fair comparison, we used the same sets of training and test data, and the same value for the feature dimension $k$.

FCNN is a network composed of neurons that connect before and after each layer, as shown in Figure~\ref{fig:FCNN}. The network learns the features by propagating information from the input layer to the output layer.
In the FCNN experiment, image data were used as inputs.
During the network learning procedure, the input data are transformed into a hidden layer, which is then activated using the ReLU function. The FCNN performs a random 20\% dropout to prevent overfitting, and the hidden layer is finally converted to the output layer, which is activated by the softmax function, similar to the NBMF. The objective function is a multiclass cross-entropy loss of the output layer and the true class information. We aimed to obtain a model with high classification accuracy by minimizing the objective function by updating the parameters, and we used Adam\cite{Kingma_2014} as a parameter optimization algorithm.
By minimizing the difference between the output of FCNN network and the ground-truth class information, the network can effectively acquire discriminative features that are conducive to image classification.

NMF performs matrix factorization\cite{Fevotte_2010,Chichoki_2009} similar to NBMF, but ensures that all components are nonnegative. However, unlike NBMF, NMF does not impose binary constraints on the coefficient matrix. In other words, the components of matrix $H$ are positive real values rather than binary values. The objective function to be minimized for both methods is the difference between matrices $V$ and $WH$. While NBMF uses combinatorial optimization for updating, NMF uses a multiplicative method and incorporates the coordinate descent method, which only computes the locations that significantly affect the components of the updated matrix.

\subsection*{Experimental settings for evaluating the performance of multiclass classification.}

To evaluate the performance of NBMF, we trained a classification model using common image data and compared it with conventional classical machine-learning algorithms.
We used the MNIST dataset, which comprises grayscale handwritten digit images.
We chose the MNIST dataset for this experiment because of the constraints imposed by the limited number of qubits in the D-Wave machine.
If we use RGB images, the data volume would surpass that of grayscale images, resulting in a highly complicated classification task.
This would require a substantial increase in feature dimensions to achieve precise classification results.
Nonetheless, the quantum annealing solver in the D-Wave 2000Q machine permitted a maximum of $k = 64$ feature dimensions.
However, stability issues arise when $k$ exceeds 40.
Thus, we plan to work on RGB datasets as an advanced task in the future.
The MNIST dataset contains 10 classes and 60,000 images. For our experiment, we randomly selected the training and test image sets from this dataset.
The number of test data images was fixed at $M = 500$, and the number of training data images, $m$, was in an arbitrary range.
The image size of MNIST is $28 \times 28$, and we concatenate the 10-dimensional class information with the image to create an input matrix $V$ of size $794 \times m$.
The coefficient of the one-hot vector is fixed at $g = 9$.
The parameters in the projected RMSProp are fixed at $\beta = 0.99$, $\eta = 0.01$, and $\epsilon = 1.0 \times 10^{-7}$.

In NBMF, the number of training epochs corresponds to the sequential updates of the matrices and is fixed at ten.
As NBMF converges quickly, we set a small number of epochs to shorten the learning time\cite{Asaoka_2020}.
For comparison, the number of epochs for the other machine-learning methods was fixed at ten.
However, in classical machine learning, the number of epochs is usually set as large as possible to achieve full convergence.
Therefore, the learning performance of classical machine-learning methods may not be fully demonstrated under only ten epochs.
In Figure~\ref{fig:epoch}, we compare the learning performances of classical machine-learning methods and NBMF when a large number of epochs are used.

We performed three sets of ten epochs for each model.
Each set begins training with random initial component values.
We recorded the highest accuracy for each set, and plotted the average of the three sets as the classification performance of the model, as shown in Figures~\ref{fig:data}a, \ref{fig:k}a, and \ref{fig:epoch}a.
Additionally, we calculated the multiclass cross-entropy loss at the final epoch of each set and plotted the average loss of the three sets as the classification performance of the model, as shown in Figures~\ref{fig:data}b and \ref{fig:k}b.
The average losses of the three sets are plotted in Figure~\ref{fig:epoch}b.

\subsection*{Annealing methods for comparison.}

The coefficient matrix $H$ in NBMF is optimized using an annealing solver for combinatorial optimization. During training, we used three different solvers.

The first solver, the D-Wave machine (DW\_2000Q\_VFYC\_6), is a quantum annealing machine equipped with qubits that uses quantum fluctuations for optimization.
Our method used a D-Wave solver to update the variables in Eq.~\mbox{\eqref{eq:H_qubo}}.
To minimize the loss function using quantum annealing via the D-Wave solver, the variables in vector $\bm{q}$ are embedded into the qubits in the machine.
It is necessary to embed these variables in a complete graph of qubits.
An automated function implemented in the D-Wave software development kit was used for embedding.
The parameters in Eq.~\mbox{\eqref{eq:H_qubo}} are the coefficients of variables $q_i$ representing the strength of the local magnetic field, and the coefficients of variables $q_iq_j$ represent the interaction parameters between the two variables.
The D-Wave solver received these parameters in a cloud environment and performed quantum annealing.
We set the number of readouts of the annealing results to 50 because there was no improvement in the performance for larger readouts.
The solution from the D-Wave solver is a binary combination of the components in vector $\bm{q}$, which corresponds to the lowest energy of the objective function among the 50 annealings.

The second solver, Fixstars Amplify Annealing Engine (AE), is an annealing machine that performs SA in parallel with GPU architecture.
Because AE is also a cloud-based solver, the procedure to minimize Eq.~\mbox{\eqref{eq:H_qubo}} obtained by SA using the AE solver was mostly the same as that obtained by quantum annealing using the D-Wave solver.
The difference between simulated and quantum annealing is that the former uses thermal fluctuations instead of quantum fluctuations.
Thermal fluctuation is controlled by a parameter corresponding to the temperature.
In our experiment, the solver automatically determined the annealing schedule and temperature values, and the duration for reading out annealing results was set to 1000 ms.
The AE solver can handle a larger number of variables simultaneously than the D-Wave solver.

Finally, for the third solver, SA, we used a package called PyQubo\mbox{\cite{Zaman_2021, Tanahashi_2019}}.
The number of reading out annealing results was set to 50, which was the same as in the D-Wave experiment.
The annealing schedule and temperature were adjusted automatically by using the SA solver.

During the test classification of NBMF, we used only SA because communication and queuing times can occur when using the D-Wave machine or AE to optimize the 500 test data. Therefore, we used only the local environment during testing to avoid waiting times.

\subsection*{Experimental settings for measuring optimization time.}

The computing speeds of the solvers were compared by minimizing the difference between the left and right sides of Eq.~\eqref{eq:v}. We generated $V$ from the product of the randomly assigned $W$ and $H$ matrices, where the dimensions of $W$ and $H$ were fixed at $n = 794$ and $k = 40$, respectively, similar to the MNIST classification experiment. The number of column vectors in $V$ and $H$ was fixed at ten and 100, respectively. We optimized $H$ by minimizing Eq.~\eqref{eq:H_qubo} and measured the average time required to optimize each column. When using the D-Wave machine and AE, communication and queuing with cloud-based machines occurred. We measured both the total execution time of the program and the annealing time, which is the original computing time for the cloud-based annealing solvers.

\section*{Data availability}

The data that support the findings of this study are available from the corresponding author upon reasonable request.
 
\bibliography{sample}

\section*{Acknowledgments}

This study was partially supported by JSPS KAKENHI Grant Number JP23H04499.
The authors thank Sigma-i Co. for their kind support.
The authors thank Editage (www.editage.jp) for English language editing.

\section*{Author contributions statement}

H.A. conceived and conducted the experiments and analyzed the results.
K.K. supervised the project.
All the authors wrote and reviewed the manuscript.

\section*{Additional information}

%To include, in this order: \textbf{Accession codes} (where applicable); \textbf{Competing interests} (mandatory statement).

\textbf{Competing interests}

The authors declare no competing interests.

%The corresponding author is responsible for submitting a \href{http://www.nature.com/srep/policies/index.html#competing}{competing interests statement} on behalf of all authors of the paper. This statement has been included in the revised manuscript.

\end{document}